\documentclass[journal,twoside,web]{ieeecolor}
\usepackage{tmi}
\usepackage{cite}
\usepackage{amsmath,amssymb,amsfonts}
\usepackage{algorithmic}
\usepackage{graphicx}
\usepackage{textcomp}

\usepackage{booktabs} 
\usepackage{hyperref}
\usepackage{algorithm}
\usepackage{algorithmic}
\usepackage{amsmath}
\usepackage{amssymb}
\usepackage{mathtools}

\usepackage{amsthm}

\usepackage{multirow} 

\usepackage[capitalize,noabbrev]{cleveref}

\usepackage{xcolor}

\usepackage{float}

\theoremstyle{plain}

\theoremstyle{definition}

\theoremstyle{remark}

\usepackage[textsize=tiny]{todonotes}

\begin{document}
\title{MRPD: Undersampled MRI Reconstruction by Prompting a Large Latent Diffusion Model}
\author{Ziqi Gao, \IEEEmembership{Student Member, IEEE}, S. Kevin Zhou, \IEEEmembership{Fellow, IEEE}
\thanks{Corresponding author: S. Kevin Zhou (e-mail: skevinzhou@ustc.edu.cn). Z. Gao and S.K. Zhou are with the School of Biomedical Engineering, Division of Life Sciences and Medicine, University of Science and Technology of China (USTC), Hefei, Anhui, China and Center for Medical Imaging, Robotics, Analytic Computing \& Learning (MIRACLE), Suzhou Institute for Advance Research, USTC, Suzhou, Jiangsu, China. Zhou is also affiliated with Key Laboratory of Precision and Intelligent Chemistry, USTC, Hefei Anhui, 230026, China and Key Laboratory of Intelligent Information Processing of Chinese Academy of Sciences (CAS), Institute of Computing Technology, CAS, Beijing, 100083, China.}
}


\maketitle
\definecolor{unsupervisedcolor}{rgb}{0.871, 0.922, 0.969}
\definecolor{supervisedcolor}{rgb}{0.984, 0.898, 0.839}
\definecolor{diffusioncolor}{rgb}{1, 0.949, 0.8}
\definecolor{mplgreen}{rgb}{0, 0.5, 0}

\begin{abstract}
Implicit visual knowledge in a large latent diffusion model (LLDM) pre-trained on natural images is rich and hypothetically universal to natural and medical images. To test this hypothesis from a practical perspective, we propose a novel framework for undersampled \textbf{MR}I Reconstruction by \textbf{P}rompting a large latent \textbf{D}iffusion model (MRPD). While the existing methods trained on MRI datasets are typically of limited generalizability toward diverse data acquisition scenarios, MRPD supports unsupervised and universally adaptive MRI reconstruction. For unsupervised reconstruction, MRSampler guides LLDM with a random-phase-modulated hard-to-soft control. With any single- or multiple-source MRI dataset, MRPD's performance is boosted universally by a lightweight MRAdapter that only finetunes the LLDM's autoencoder. Experiments on FastMRI and IXI show that MRPD is the only model that supports both MRI database-free and database-available scenarios and attains the best generalizability towards out-of-domain (OOD) samplings, contrasts, and organs among compared unsupervised, supervised, and MRI diffusion methods. To our knowledge, MRPD is the first method that empirically shows the universal prowess of an LLDM pre-trained on vast natural images for MRI. Our official implementation is at https://github.com/Z7Gao/MRPD. 
\end{abstract}


\section{Introduction}
\label{sec:introduction}
Magnetic Resonance Imaging (MRI) is a cornerstone in clinical diagnostics renowned for its safety. Diverse high-contrast MRI, including T$_1$-weighted (T$_1$-w), T$_2$-weighted (T$_2$-w), and Proton Density-weighted (PD-w) images, enables comprehensive radiological assessments. However, MRI's slow acquisition speed limits its clinical efficiency, due to inherent physical and physiological constraints\cite{CSlustig2008compressed}.
To this end, accelerated MRI techniques with undersampling acquisitions have been developed, the most notable ones including compressive sensing (CS) \cite{CSlustig2008compressed,CShaldar2011random,CSpatel2011gradient,CSliang2009accelerating,2007CSMRI_TV}, parallel imaging (PI) \cite{PIgriswold2002GRAPPA,PIpruessmann1999SENSE}, and a recent trend, deep learning (DL) \cite{wang2016unet,jin2017unetmri,zhu2018nature,qin2018dc1,schlemper2017dc,eo2018kiki,putzky2019rim,cheng2019spd,huang2022swinmr,yang2017dagan,quan2018compressedgan}.
\begin{figure}[t]
    \centering
    \includegraphics[width=0.9\columnwidth]{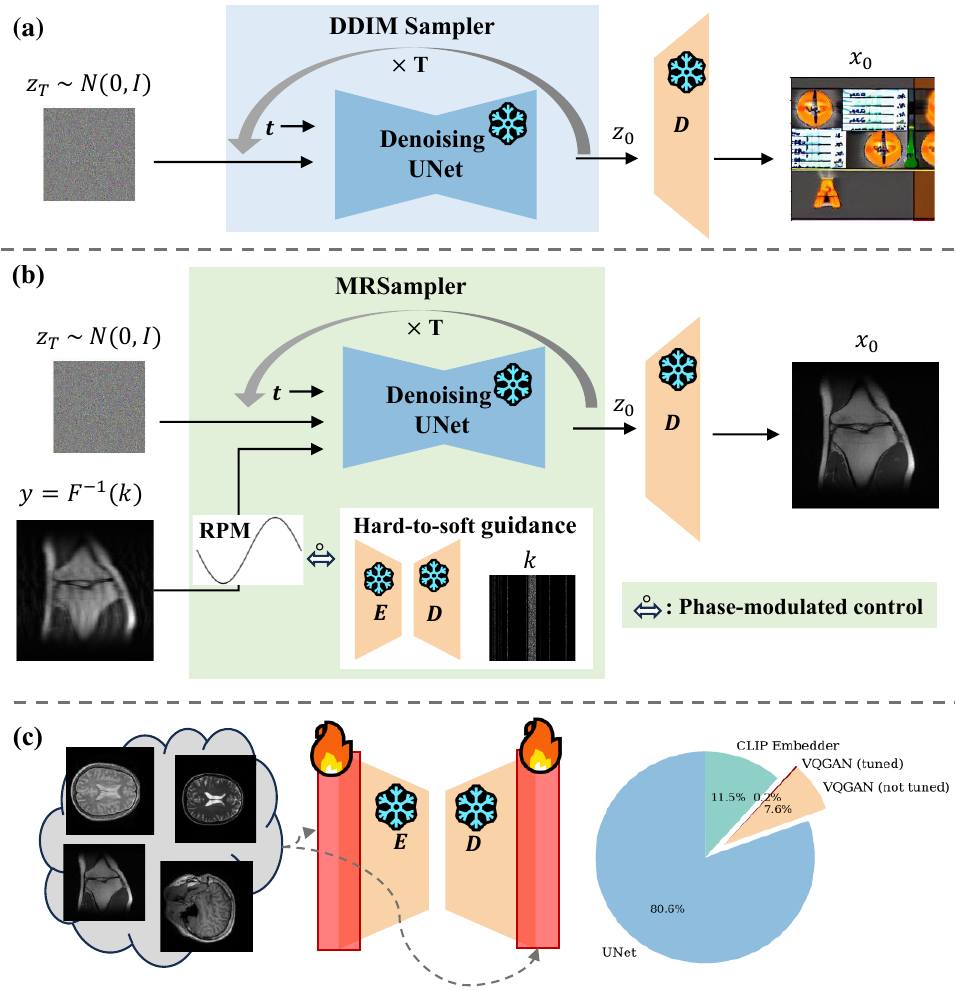}
    \caption{Overview of MRPD. (a) Unconditional image generation of LLDM with a deterministic DDIM sampler. (b) Image-specific undersampled MRI reconstruction with an MRSampler. (c) MRAdapter for MRI database-available scenarios and the composition of an LLDM, Stable Diffusion v1.5, with our adaptation.} \vspace{-5mm}
    \label{fig:mainDiagram}
\end{figure}


Recently, MRI diffusion models (DM)\cite{song2021scoresde,peng2022DiffuseRecon,chung2022scoreMRI,nips2021score,GUNGOR2023AdaDiff,ozturkler2023smrd,Cao2024HFS-SDE,korkmaz2023SSDiffRecon,huang2023cdiffmr,ozturkler2023red-diff} have emerged as competitive to state-of-the-art supervised methods\cite{zhou2020dudornet,Guo2023ReconFormer}. Instead of using a discriminative objective depending on particular sampling patterns, they train a diffusion model to approximate the distribution of MR images in a sampling-agnostic way. Score-MRI\cite{chung2022scoreMRI} has demonstrated its notable generalization capability across diverse organs and modalities. However, the dataset scale for training the current MRI diffusion models \cite{song2021scoresde,chung2022scoreMRI}  typically ranges from thousands to tens of thousands of samples and it remains challenging to curate large-scale MR images due to privacy issues\cite{wang2023onegeneralizable}. Parallel to this, large latent diffusion models (LLDM) \cite{podell2023sdxl,rombach2021ldm}, a class of revolutionary generative foundation models exemplified by Stable Diffusion (SD), are trained on vast Internet-scale images \cite{schuhmann2022laion5b} of about $10^{6}$ times larger than existing public MRI datasets \cite{fastmri1,fastmri2} used for training MRI diffusion models. SD models can generate diverse, high-fidelity images and have proven a powerful backbone \cite{zhao2023unleashingvp, xu2022odise_open_vocab, yang2023diffusionsurvey,ke2024cvprDepthSD} with supervised fine-tuning. These facts imply the richness of visual knowledge in an LLDM and prompt our questions: \textit{(Q1) Is the implicit visual knowledge in the LLDM pre-trained on natural images \underline{universal for both natural and medical images}? (Q2) If yes, how can it be harnessed and enhance the \underline{adaptability} and \underline{generalizability} of undersampled MRI reconstruction models?} Given LLDMs' broad applicability, we hypothesize that universal visual knowledge exists and can be harnessed for MRI data. To answer the first question from a practical perspective, we provide a solution, MRPD, to the second question.

 As the most related area of research in computer vision, Diffusion model-based Inverse problem Solvers (DIS) perform conditional posterior sampling using the pre-trained diffusion prior. The majority of DIS focus on pixel \cite{DIS1,2022DDRM,wang2023ddnm,DIS2, chung2023DPS,chung2022MCG,pmlr-v202-song23kLGD-MC,song2023pseudoinverseGDM} while few works explore their effectiveness in the latent domain\cite{rout2023psld,song2023resample,chung2023text}. Furthermore, most of the research concentrates on pixel-domain solvers using in-distribution data and has not explored complex-valued image restoration. In this context, we investigate the challenging combination of performing \textbf{latent} diffusion model sampling for \textbf{extreme out-of-distribution} (OOD) and \textbf{complex-valued} data.  

In this work, we propose an unsupervised and universally adaptive undersampled MRI reconstruction framework called undersampled \textbf{M}RI \textbf{R}econstruction by \textbf{P}rompting a pre-trained latent \textbf{D}iffusion model (MRPD). By prompting an LLDM pre-trained on billions of natural images, MRPD recovers the fully-sampled image from the undersampled k-space data in an unsupervised and universally adaptive way. Specifically, our MRSampler, tailored for complex-valued MRI images, allows the training-free diffusion inference with novel hard-to-soft consistency guidance and random phase modulation. Our MRAdapter is an efficient LDM fine-tuning technique that only refines the data interface of an LLDM, the autoencoder, instead of the large denoising UNet. Given any single- or multi-source MRI data, our MRAdapter can improve the reconstruction ability universally and preserve the generalizability. We evaluate MRPD on both real simulation datasets and complex single-coil and multi-coil datasets. Our advantage over other existing DL methods is listed in Table \ref{table:summary} and the key contributions are as follows:


\begin{table}[H]
    \vspace{-2mm}
    \centering
    \caption{Characteristics of DL models for MRI reconstruction.}
    \resizebox{\columnwidth}{!}{
    \begin{tabular}{lcccc}
    \hline
     Characteristics \textbackslash \ Model type & \textbf{Supervised models} & \textbf{INR} & \textbf{Diffusion} & \textbf{MRPD} \\ \hline
    \textbf{Sampling-agnostic models.} & $\times$  & \checkmark & \checkmark & \checkmark \\
    \textbf{Utility of large-scale data.} & \checkmark & $\times$  & \checkmark & \checkmark \\
    \textbf{Adaptability to database-free scenarios.} & $\times$  & \checkmark & $\times$ & \checkmark \\
    \textbf{Generalizability to OOD organs and contrasts.} & Typically poor & \checkmark & Average & \checkmark \\
    \hline
    \end{tabular} 
    }
    \vspace{-1mm}
    \label{table:summary}
\end{table}

\begin{itemize}
    \item To our knowledge, MRPD are {\bf the first} to successfully retrieve the implicit visual knowledge in an LLDM pre-trained on vast natural images for medical imaging inverse problems, particularly accelerated MRI.
    \item MRPD uses an MRSampler for guiding LDMs in an unsupervised way. Optionally, an efficient LDM autoencoder adapter, MRAdapter enables utilizing any available MRI dataset in a universal and sampling-agnostic way.
    \item Extensive evaluations under various in-domain (In-D) and OOD conditions including samplings, organs, and contrasts show that MRPD achieves prominent improvements in both adaptability and generalizability over popular unsupervised, supervised, and MRI diffusion methods. 
\end{itemize}

\section{Background and Related Work}
\subsection{Diffusion Models}
Diffusion model (DM)\cite{ho2020ddpm,Dhariwal2021ImprovedDDPM,song2021scoresde} is a class of unconditional generative methods that transform a Gaussian distribution into an empirical data distribution. The forward process is a Markovian process that gradually adds Gaussian noise with a fixed schedule to a clean image $x_0$ and gets intermediate noisy $x_t$ in the $t$ step. The reverse process involves using a pre-trained neural network, i.e. \textit{denoising function},  to gradually denoise $x_t$. To reduce the computational burden, the latent diffusion model (LDM)\cite{rombach2021ldm} is proposed. It embeds data in a lower-dimension latent space before learning the denoising function. LLDM\cite{rombach2021ldm,dalle, saharia2022imagen} is a class of LDM trained on Internet-scale dataset.  We demonstrate the composition of Stable Diffusion (SD)\cite{rombach2021ldm} v1.5 in Fig.~\ref{fig:mainDiagram}. It has CLIP \cite{clip} as its text encoder, trains an autoencoder, VQGAN\cite{esser2020tamingVQGAN}, and a latent image generator, denoising UNet, on the LAION-5B dataset\cite{schuhmann2022laion5b} consisting of 5.85M natural image-text pairs. The implicit visual knowledge lies in the denoising UNet.  

Given a pre-trained denoising function, diffusion samplers are designed to tailor specific needs while keeping the intermediate images on the correct noisy data manifold. Otherwise, the generated image will suffer from strong aliasing\cite{chung2022MCG}. Denoising diffusion implicit model (DDIM) sampler\cite{song2021DDIM} is the backbone of MRPD. It generates high-quality samples with few reverse steps, ideal for time-sensitive inverse problems. The diffusion process is defined as a non-Markovian process: in each step $t$, the clean latent image \(\hat{z}_0\) is predicted via 
\begin{equation}
\hat{z}_{0} = \sqrt{\alpha_{t-1}} \left( \frac{z_t - \sqrt{1 - \alpha_t} \cdot \epsilon_{\theta}^{(t)}(z_t)}{\sqrt{\alpha_t}} \right),
\label{equation: latent deterministic DDIM sampling predict}
\end{equation}
, where \(\epsilon_{\theta}^{(t)}\) is the pre-trained denoising function and $\alpha_t$ denotes the noise schedule. Then it samples \(z_t\) from \(z_{t+1}\) via
\begin{equation}
z_{t-1} = \hat{z}_{0}+ \sqrt{1 - \alpha_{t-1} - \sigma_t^2} \cdot \epsilon_{\theta}^{(t)}(z_t).
\label{equation: latent deterministic DDIM sampling}
\end{equation}
\(\sigma_t\), the standard deviation of Gaussian noise at time \(t\), controls the stochasticity of the diffusion process. We use a deterministic DDIM sampler for LLDMs (\(\sigma_t=0\)).

\subsection{Diffusion Models for MRI Reconstruction}
Pioneering works \cite{song2021scoresde,peng2022DiffuseRecon,chung2022scoreMRI} introduce diffusion models for undersampled MRI reconstruction and report high-quality and sampling-agnostic reconstruction\cite{chung2022scoreMRI} performance. Recent works also explore diffusion-based MRI reconstruction models in the domain of self-supervised learning\cite{korkmaz2023SSDiffRecon} and inference acceleration\cite{ozturkler2023red-diff,chung2022ccdf,Cao2024HFS-SDE}. Two notable papers\cite{GUNGOR2023AdaDiff,ozturkler2023smrd} aim at mitigating distribution shifts of pre-trained diffusion models. AdaDiff\cite{GUNGOR2023AdaDiff} fine-tunes a pre-trained rapid adversarial diffusion model with data consistency loss. However, it is built upon a particular type of adversarial DMs instead of a regular diffusion model\cite{rombach2021ldm} commonly used for large-scale foundation model training. SMRD\cite{ozturkler2023smrd} proposes a SURE-based test-time hyperparameter selection technique for sampling a CSMG framework\cite{nips2021score}, a pixel-DIS instead of latent-DIS. None of the existing diffusion MRI models are latent diffusion models and no successful attempt in utilizing LLDMs, or general generative foundation models, emerged.

\section{Method}
\subsection{Problem formulation}
We consider the following forward measurement model for accelerated MRI: 
\begin{equation}
    k= M\mathcal{F}\mathcal{S}x_{f}+\epsilon, y=\mathcal{F}^{-1}k,
\end{equation}
where $x_f \in \mathbb{C}^n$ is the original fully-sampled image, $k\in\mathbb{C}^n$ is the k-space measurement per coil, $y\in\mathbb{C}^n$ is the undersampled (aliased) image per coil, and $\epsilon$ is the measurement noise. $M \in \{0,1\}^n$ is the undersampling mask and $(\mathcal{F}, \mathcal{F}^{-1})$ is the Fourier Transform pair.  $\mathcal{S}:=[S^{(1)}; S^{(2)}; ...; S^{(c)}]$ is the sensitivity map for $c$ coils. Undersampled MRI reconstruction aims at recovering high quality $\hat{x}$ from undersampled $k$:
\begin{equation}
    \hat{x} = \mathop{\arg\min}\limits_{x} ||k-M\mathcal{F}\mathcal{S}x|| + \lambda \cdot \mathcal{R}(x),
    \label{formulation}
\end{equation}
where the first term enforces the fidelity of k-space measurement and $\mathcal{R}$ is some image regularizers. For the single-coil acquisition, $\mathcal{S}$ is reduced to an identity matrix.

MRPD utilizes the universal image prior from an LLDM as $\mathcal{R}$ and incorporates the k-space fidelity term in the denoising process progressively by MRSampler (sec. \ref{sec:MRSampler}). Given any MRI data, MRAdapter improves the accuracy of the data fidelity term in a universal and undersampling-agnostic way(sec. \ref{subsec: MRAdapter}). For simplicity, we first introduce our main components in the single-coil case and then describe our adaptability towards multi-coil acquisitions (sec. \ref{sec:PI}). 

\subsection{Unsupervised MRSampler}
\label{sec:MRSampler}
\begin{algorithm}[h]
    \caption{MRSampler}
    \label{alg:mrsampler}
    \textbf{Input:} $k$, $M$. \\
    \textbf{Require:} an LLDM (autoencoder: \(\mathcal{E}\) and \(\mathcal{D}\)), a deterministic DDIM Sampler \(\mathcal{S}\) with a total timestep $T$ (default to 1000). \\
    \textbf{Hyperparameters:} $t_0$, $t_{ws}$, $\lambda$, soft CG scale $\gamma$.  

    \begin{algorithmic}[1]
        \STATE $y = \mathcal{F}^{-1}(k)$ \hfill \textcolor{gray}{// the undersampled image}
        \STATE $\theta_y = arg(y),\theta_r \sim \mathcal{U}(-\pi,\pi)$ 
        \STATE $\hat{\theta} = \lambda\theta_r+(1-\lambda)\theta_y$
        \hfill \textcolor{gray}{// the modulated image phase}
        \STATE $k^{RPM}= \mathcal{F}(|y|e^{j\hat{\theta}})$  
        \STATE $\epsilon \sim \mathcal{N}(0, I)$ \hfill \textcolor{gray}{//}
        \STATE $z_{t_0\cdot T} = \sqrt{\alpha_{t_0}}\mathcal{E}(|y|)+\sqrt{1-\alpha_{t_0}}\epsilon$  
        \FOR{$t=t_0 \cdot T-1$ to $0$}
            \STATE $\hat{z}_0 = \mathcal{S}.\text{predict}(z_{t+1},t+1)$  \hfill \textcolor{gray}{// Eq.(\ref{equation: latent deterministic DDIM sampling predict})}
            \STATE $\hat{x}_0 = \mathcal{D}(\hat{z}_0)$
            \STATE ${\hat{k}}^{RPM} = \mathcal{F}(\hat{x}_0e^{j\hat{\theta}})$
            \IF{$t > t_{ws} \cdot T$}
                \STATE \textcolor{gray}{// hard DC}
                \STATE $\hat{k}^{dc} = (1-M) \cdot {\hat{k}}^{RPM}+M \cdot k^{RPM}$
                \STATE $\hat{x}^{dc} = \mathcal{F}^{-1}(\hat{k}^{dc})$, $\hat{z}_0^{dc}=\mathcal{E}\{|\hat{x}^{dc}|\}$
                \STATE $z_t = \mathcal{S}.\text{sample}(z_{t+1}, t, \hat{z}_0^{dc}, \text{cg}=0)$  \hfill\textcolor{gray}{// Eq.(\ref{equation: latent deterministic DDIM sampling})}
            \ELSE
                \STATE \textcolor{gray}{// soft CG}
                \STATE $z_t = \mathcal{S}.\text{sample}(z_{t+1}, t, \hat{z}_0, \text{cg}=-\gamma \cdot \nabla_{z_{t+1}}\|M \cdot k^{RPM}-M \cdot {\hat{k}}^{RPM}\|)$  
            \ENDIF
        \ENDFOR
        \STATE $x_0 = \mathcal{D}(z_0)$ 
        \STATE return $x_0$
    \end{algorithmic}
\end{algorithm}
\vspace{-0.5mm}
MRSampler is a physically consistent LDM solver for MRI. It supports generalizable undersampled MRI reconstruction in both MRI database-free and database-available scenarios. It builds upon a deterministic DDIM sampler\cite{song2021DDIM}, with several novel phase-modulated guidance mechanisms to enhance its efficacy for complex-valued MRI. Algorithm \ref{alg:mrsampler} provides a formal description of the MRSampler.

\subsubsection{Hard-to-Soft Diffusion Guidance} 
Given the k-space measurement $k$, MRSampler uses hard-to-soft diffusion guidance to ensure the generated images are consistent with $k$. At each denoising step, hard data consistency (DC) imposes stringent measurement constraints by replacing the generated unmasked k-space with the measurements and soft classifier guidance (CG) uses the gradient given by the measurement constraints to modify the results given by $\epsilon_{\theta}$ (Eq. \ref{equation: latent deterministic DDIM sampling}) \cite{Dhariwal2021ImprovedDDPM}; yet none of the above-mentioned unitary guidance of latent-DIS has demonstrated stability with extreme OOD data. Hard DC (Fig. \ref{ResultOfRSampler}(c)) leads to stronger artifacts, demonstrating that the intermediate images are away from the correct noisy data manifold; soft CG (Fig. \ref{ResultOfRSampler}(d)) generates alias-free but overly smooth results. 

Interestingly, the complementary behavior of hard DC and soft CG is observed from intermediate results of Fig. \ref{ResultOfRSampler}: hard DC generates a rough draft within only a few steps, while soft CG generates decent images at last but much slower than hard DC in the first steps. As such, we propose the composite hard-to-soft diffusion guidance: the first few steps leverage the fast convergence of hard DC while soft CG is used for later steps to refine the image within the correct noisy data manifold. During reconstruction, MRSampler enforces either hard DC (Algorithm \ref{ResultOfRSampler}, line 11-15) or soft CG (Algorithm \ref{ResultOfRSampler}, line 16-18) depending on the current timestep $t$ and a hyperparameter, watershed timestep $t_{ws}$, after attaining the current clean latent prediction $\hat{z}_0$ and the pixel-domain one $\hat{x}_0=\mathcal{D}(\hat{z}_0)$. To speed up, the early steps are skipped and the sampling starts at $t_0\cdot T$, while the initial $|y|$ perturbed with a scheduled noise is sent to the LLDM.

\subsubsection{Random Phase Modulation}
\begin{figure}
    \centering
    \includegraphics[width=0.45\textwidth]{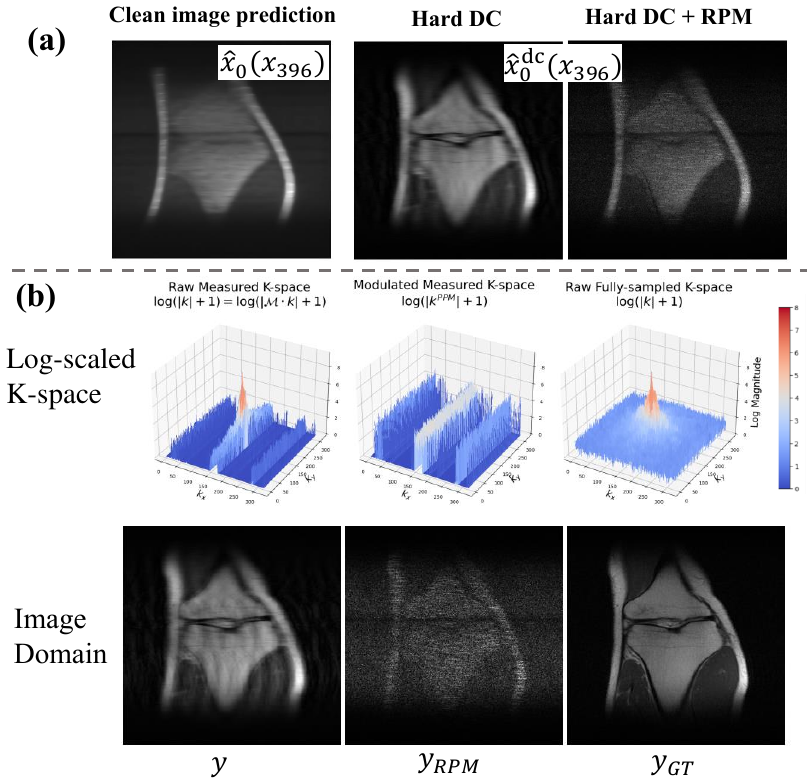}
    \caption{RPM's effect. Panel (a) shows an intermediate clean image $\hat{x}_0$ guided by hard DC without and with RPM and (b) shows the impact of RPM on the k-space domain with the corresponding images.}
    \label{fig:PPM}
    \vspace{-0.15in}
\end{figure}

We observe undesired results when solely applying the hard-to-soft diffusion guidance. The hard DC step can push the diffusion inference path out of the manifolds\cite{chung2022MCG}, leading to inaccurate reconstruction, for example, the strong artifact shown in Fig. \ref{ResultOfRSampler}(d). The soft CG is based on the gradient of the measurement consistency loss on the k-space domain, which has particular distribution characteristics across frequencies\cite{Mezrich1995APOkspace}. As shown in Figure \ref{fig:PPM}(b), the raw fully-sampled k-space has much larger low-frequency components than high-frequency ones. As a result, the loss is biased toward low-frequency components, leading to overly smooth reconstructed images shown in Fig. \ref{ResultOfRSampler}(c).

We tackle the two issues with a simple but effective random phase modulation (RPM) operation for joint noise injection and frequency normalization. Given the measurement, MRSampler first transforms $k$ into an under-sampled complex-valued image $y=\mathcal{F}^{-1}(k)$. Then, the original image phase $\theta_y = arg(y)$ is combined with a random image phase, $\theta_r \in [-\pi,\pi)$ with a random coefficient $\lambda \in [0,1]$. The combined image phase $\hat{\theta} = \lambda\theta_r+(1-\lambda)\theta_y$ replaces $\theta_y$ and the original k-space measurement is modulated into $k^{RPM}= \mathcal{F}(|y|e^{j\hat{\theta}})$. During sampling, the predicted magnitude-only image $\hat{x}_0$ is combined with $\hat{\theta}$ before transforming into the predicted modulated k-space, $\hat{k}^{RPM}$. Hard DC and soft CG steps enforce the data consistency between $k^{RPM}$ and $\hat{k}^{RPM}$.

Fig. \ref{fig:PPM} illustrate RPM's influence on hard DC and soft CG. 
Fig. \ref{fig:PPM}(a) shows the clean prediction $\hat{x}_0$ and consistent reconstruction $\hat{x}_0^{dc}$ at step 396, a very early step given the sampling starts at 400 ($t_0=0.4$). An intermediate clean image $\hat{x}_0(x_{396})$ is predicted and the right two images are the data-consistent reconstruction $\hat{x}_0^{dc}(x_{396})$ given by hard DC and hard DC + RPM. Hard DC mitigates the uncertainty of the predicted clean image but introduces global aliasing; RPM injects a noise globally that makes the OOD  artifacts unrecognizable, effectively keeping the diffusion inference path within the correct manifolds. Fig. \ref{fig:PPM}(b) shows the effect of RPM in the k-space domain and the corresponding images. With RPM, the distribution among frequencies is more balanced than the raw distribution. As such, soft CG concentrates less on the dominant low frequencies and more on high-frequency image detail reconstruction (Fig. \ref{fig:main}, \textcolor{yellow}{yellow} and \textcolor{red}{red} boxes).


\subsection{Universally Adaptive MRAdapter}
\label{subsec:MRAdapter}
Instead of proposing a static unsupervised model, we discover that the LLDM allows for building an adaptive model for undersampled MRI reconstruction so that any given MRI data can improve the performance universally. A straightforward way is to use existing SoTA LLDM finetuning techniques, e.g. ControlNet\cite{zhang2023controlnet}, to adjust the denoising UNet. However, adapting the data prior is both computationally inefficient and inevitably harmful to LLDM's generalizability. Instead, we observe that both the pixel-domain soft and hard constraints of MRSampler pass through the autoencoder, a lossy image compressor that leads to information loss. Therefore, we propose an MRAdapter that adapts part of the autoencoder, accounting for merely 0.2\% of the entire LLDM's parameters (Fig. \ref{fig:mainDiagram}(c)). Specifically, given any MR image, we finetune the first and last layers of the autoencoder with an L$_2$ loss. The MRI image magnitude is retrieved, normalized to [-1,1], and duplicated three times to align with the autoencoder interface. The convolutions become two sets of image filters: MR-to-natural-image at the start for adapting the input to LLDM and natural-to-MR-image at the end for refining the MRI output. MRAdapter improves MRI reconstruction while preserving the generalizability of generative foundation models. With any single- or multi-source MRI dataset, MRAdapter improves reconstruction universally, discussed and evaluated in Sec. \ref{sec:universal}.

\subsection{MRPD meets Parallel Imaging (PI)}
\label{sec:PI}

Many modern MRI scanners have multiple coils\cite{fastmri1}. We propose a multi-coil variant of MRPD based on GRAPPA\cite{PIgriswold2002GRAPPA}. GRAPPA is a widely adopted PI algorithm in clinical practices. It generates a multi-coil reconstruction by taking the sum-of-root-sum-of-squares (SSOS) of each reconstructed single-coil image. Specifically, MRPD first reconstructs images coil-by-coil, as shown in Algorithm \ref{alg:mrsampler}. Then SSOS is applied on all reconstructed coils and generates a multi-coil reconstruction. With fully-sampled multi-coil data, the MRAdapter can use each coil separately for fine-tuning the LLDM.

\section{Experiments}
In this section, we aim to answer the key question: can the implicit visual knowledge in a pre-trained LLDM on natural images be harnessed and enhance the (1) adaptability and (2) generalizability of undersampled MRI reconstruction models?  

\subsection{Experimental Settings}
\subsubsection{Datasets} Experiments are performed on the IXI multi-contrast brain dataset\footnote{brain-development.org/ixi-dataset, CC BY-SA 3.0 license} and the FastMRI knee dataset\footnote{fastmri.med.nyu.edu}\cite{fastmri1,fastmri2}. IXI contains coil-combined magnitude images of T$_1$-w, T$_2$-w, and proton density-weighted (PD-w) acquisition simulated as single-coil data with 576 overlapped subjects. Volumes from all overlapped subjects are uniformly sampled slice-wise and then split volume-wise, resulting in (3218, 2455, 2455) slices for training and (912, 700, 700) slices for testing (matrix size = 256$\times$256). The FastMRI dataset includes 1,172 raw 15-coil and emulated single-coil complex-valued PD-w k-space data. We follow the pre-processing of \cite{chung2022scoreMRI}, specifically, we use the official dataset split and drop the first and last five noisy slices from each volume, resulting in 25012 training slices and 5145 testing slices (matrix size = 320$\times$320). For PI experiments, we resorted to 10 volumes for testing due to computational limitations. No slice from any IXI/FastMRI subject overlaps in the training and testing sets; 5\% slices of training sets are held out for validation.

\subsubsection{Baselines} We employ representative undersampled MRI reconstruction approaches from three categories including database-free CS, supervised DL, and diffusion methods as baselines, using their official implementation unless noted. Total Variation (TV)\cite{2007CSMRI_TV} is a representative CS method and we use the implementation from sigpy.mri\footnote{https://github.com/mikgroup/sigpy}. For supervised DL methods, we choose a representative UNet\cite{fastmri2}. We additionally compare against the performant supervised methods. For the real-valued simulation study and single-coil experiment, we compare against DuDoRNet\cite{zhou2020dudornet}. For multi-coil experiments, we use E2E-VarNet\cite{sriram2020e2evar}. For representative MRI diffusion methods, we choose a score-based model, Score-MRI\cite{chung2022scoreMRI}, and a denoising diffusion model, DiffuseRecon\cite{peng2022DiffuseRecon}. We compare with Score-MRI's real-type, SENSE-type, and SSOS-type algorithms for real, single-coil complex, and multi-coil complex MRI data, respectively. We also deploy CCDF to accelerate Score-MRI into 40 steps, which is reported to maintain the performance when reducing the steps from 2000\cite{chung2022scoreMRI}. DiffuseRecon is set to 200 steps.

\subsubsection{Metrics \& Sampling} To quantitatively evaluate the reconstruction results, we use peak signal-to-noise ratio (PSNR) and structural similarity index measure (SSIM). Four Cartesian sampling patterns are used in the experiments: Uniform 1D, Gaussian 1D, Gaussian 2D, and VD Poisson Disk. Uniform1D sampling is deployed unless noted. Mask generation functions are from the official FastMRI\footnote{https://github.com/facebookresearch/fastMRI} and Score-MRI\footnote{https://github.com/HJ-harry/score-MRI} repositories.

\subsubsection{Implementation Details}
\label{Implementation Details}
\colorbox{unsupervisedcolor}{Database-free methods'} hyperparameters are selected through grid-searching 100 slices from the training sets. While they can also be grid-searched on the testing set to make those methods truly database-free, we deliberately avoid this for fairness, so that no testing data is available for any method. \colorbox{supervisedcolor}{Supervised methods} are trained under uniform 1D sampling masks, with acceleration rates (R) ranging from 2x to 8x and Auto-Calibration Signal (ACS) fractions ranging from 4\% to 12.5\% for sufficient epochs: 100 for FastMRI and 1000 for IXI. They are evaluated with their best checkpoints from earlier epochs before overfitting. This training strategy improves the generalizability without reducing the in-domain performance.  \colorbox{diffusioncolor}{MRI diffusion methods} are trained or adapted on the same single-coil FastMRI PD knee training set, while DiffuseRecon uses the complex-valued data, and Score-MRI and ours only use the magnitude part.

Our method has both unsupervised (\colorbox{unsupervisedcolor}{ours.A}) and single-coil MRI-adapted (\colorbox{diffusioncolor}{ours.B}) versions. MRSampler's hyperparameters are $t_0=0.4$, $t_{ws}=0.3$, $\lambda=1$, selected by grid-searching possible values on 100 slices of the training sets (section \ref{t0,tws} and \ref{lambda}). For $t_0$ and $t_{ws}$, we choose them under a time budget of 50s per slice, which is the inference time of DiffuseRecon\cite{peng2022DiffuseRecon}. The hard DC is applied every two steps for accelerations. In all experiments, $\gamma$ is basically consistent: $\frac{1}{100}$ for FastMRI and $\frac{1}{200}$ for IXI, except for 4x multi-coil experiments ($\gamma=\frac{1}{150}$). As for the MRAdapter, we finetune the autoencoder with a learning rate = 0.0001 and an Adam optimizer ($\beta_1=0.5, \beta_2=0.9$) for 10 epochs. While crafting the initial random noise potentially improves generation \cite{Everaert_2024_WACV_noise}, we keep a fixed initial noise to control variables. 

Experiments are done with an NVIDIA GeForce RTX 3090 GPU with 32GB memory and code is implemented in PyTorch.

\subsubsection{LLDM} We use the SD v1.5 checkpoint\footnote{https://huggingface.co/runwayml/stable-diffusion-v1-5} pre-trained on billions of natural images \cite{schuhmann2022laion5b} as the pre-trained LLDM. LLDM models' data interface is through 3-channel RGB images ranging from -1 to 1 for stable training and denoising\cite{ho2020ddpm,rombach2021ldm}. Consequently, any given undersampled complex-valued k-space slice is first transformed into the image domain with inverse Fourier Transform, normalized, and duplicated the magnitude into a 3-channel input to LLDM. During sampling, the modulated phase is introduced in the data consistency step. 

\subsection{Experiment Design}
We conduct the following experiments: (1) We test the performance of ours and compared methods on the single- and multi-coil FastMRI knee dataset with various R's and Cartesian sampling patterns, including In-D and OOD ones. (2) We compare the generalizability of all methods with OOD contrasts and organ on the multi-contrast IXI brain datasets. (3) We show our method's capability of universal reconstruction by using MRAdapter to fine-tune LLDM's autoencoder with single- or multiple-source MRI datasets. We evaluate the performance of both autoencoder's and undersampled reconstruction. (4) We show our hyper-parameter selection process on FastMRI. (5) We conduct ablation studies of each component of MRPD on FastMRI to show our effectiveness. 
\section{Results}

\subsection{FastMRI Studies}
 \begin{figure*}[]
    \centering
    \includegraphics[width=0.9\textwidth]{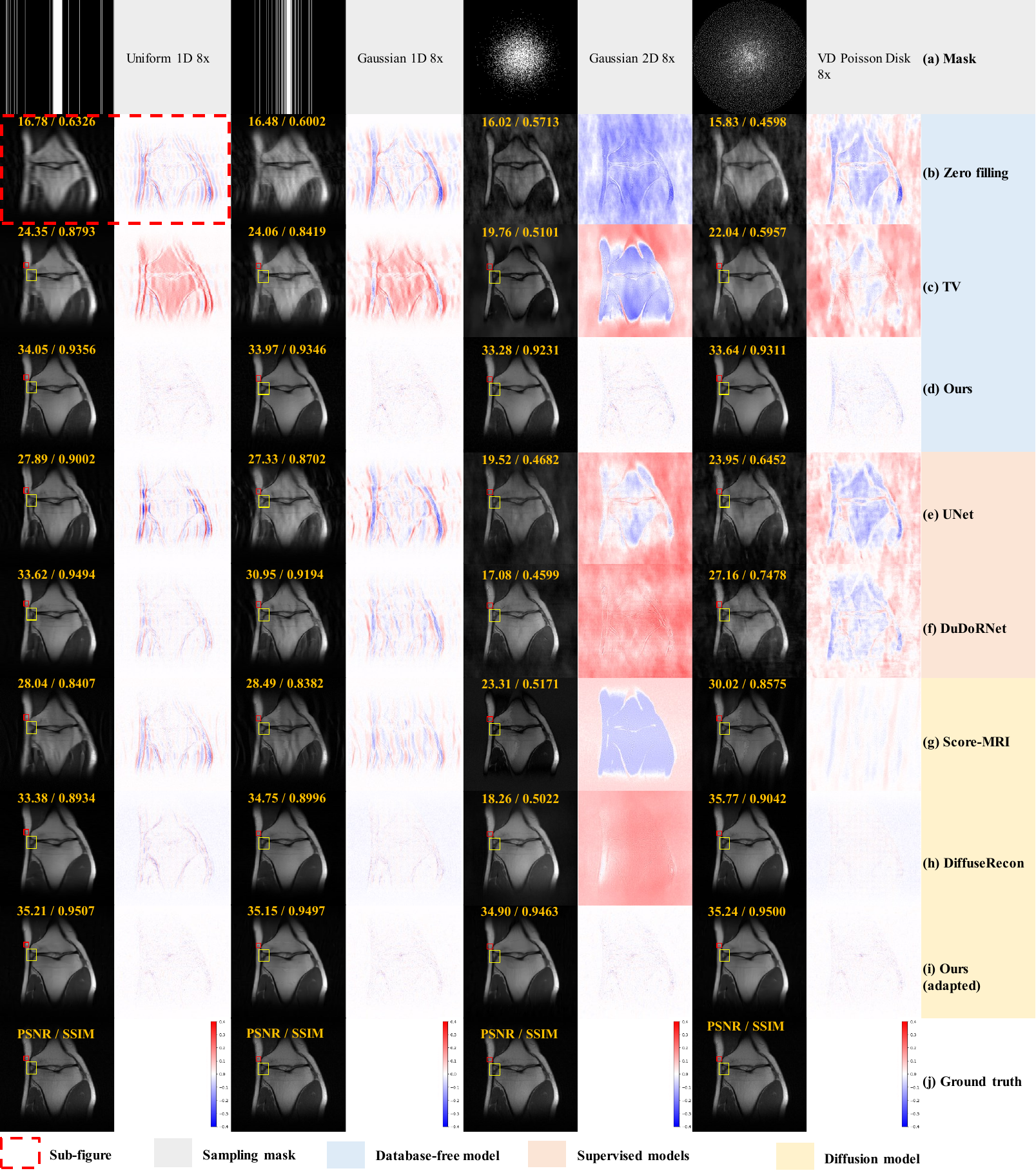}
    \caption{Visualization of an in-D PD-w knee image undersampled with diverse masks. A sub-figure has a reconstructed image and its error map.} 
    \label{fig:main} 
\end{figure*}

\begin{table*}[]
\caption{Single-coil experiments on FastMRI. PSNR (dB) and SSIM (\%) are reported as mean±std. The \textbf{best} results are marked.}
\centering
\resizebox{0.8\textwidth}{!}{
\begin{tabular}{@{}c|cc|ccccccc@{}}
\toprule
 Sampling Patterns                                 &     &      & \colorbox{unsupervisedcolor}{TV} & \colorbox{supervisedcolor}{UNet} & \colorbox{supervisedcolor}{DuDoRNet} & \colorbox{diffusioncolor}{Score-MRI} & \colorbox{diffusioncolor}{DiffuseRecon}   & \colorbox{unsupervisedcolor}{Ours.A} & \colorbox{diffusioncolor}{Ours.B} \\ \midrule
\multirow{6}{*}{Uniform 1D} & \multirow{2}{*}{4x}  & PSNR &25.35±1.95  & 30.71±2.30 & \textbf{32.94±3.09} & 30.33±2.44 & 31.38±3.74  & 30.88±2.66 & {31.50±3.03}  \\
                                  &     & SSIM &71.44±10.08 & {88.99±5.07} & \textbf{90.94±5.48} & 87.39±6.03 & 82.05±10.05  & 85.62±8.24 & 86.60±8.35 \\ \cmidrule(l){2-10} 
                                  & \multirow{2}{*}{8x} & PSNR & 24.96±1.92  & 29.18±2.26 &  {30.87±2.67} & 28.81±2.21 & 29.51±3.06  & 30.26±2.65 & \textbf{31.28±3.10} \\
                                  &     & SSIM & 64.34±14.36 & 85.63±6.53 & \textbf{87.29±7.04} & 84.23±7.01 & 76.23±11.92 & 83.97±8.41 &  {85.98±8.53} \\ \cmidrule(l){2-10} 
                                  & \multirow{2}{*}{12x} & PSNR & 24.67±1.90  & 27.56±2.17 & 28.94±2.30 & 28.24±2.07 & 28.56±2.82  &  {29.64±2.60} & \textbf{30.95±3.03}    \\
                                  &     & SSIM & 61.90±15.02 & 82.82±6.96 &  {84.41±7.44} & 83.06±7.33 & 72.65±13.12 & 82.51±8.65 & \textbf{85.41±8.86}  \\ \midrule
\multirow{4}{*}{Gaussian 1D}      & \multirow{2}{*}{4x}  & PSNR &     25.74±1.92  & 31.21±2.31  & \textbf{32.82±3.11} & 30.96±2.47 &  {31.59±3.89}  & 30.67±2.90 & 31.48±3.02    \\
                                  &     & SSIM &  71.87±11.89 &  {89.09±5.20}  & \textbf{90.51±5.69} & 88.53±6.09 & 81.93±10.49 & 85.43±8.05 & 86.55±8.29   \\ \cmidrule(l){2-10} 
                                  & \multirow{2}{*}{8x}  & PSNR & 25.07±1.99  & 28.32±2.01  & 29.97±2.36 & 28.33±2.13 & 30.29±3.55  &  {30.38±2.91} & \textbf{31.25±3.08}    \\
                                  &     & SSIM &  64.07±12.56 & 83.91±5.93  & \textbf{86.17±6.49} & 83.96±6.68 & 76.50±12.58 & 84.01±8.44 &  {85.89±8.58} \\ \midrule
\multirow{4}{*}{Gaussian 2D}      & \multirow{2}{*}{8x}  & PSNR &  25.20±2.55  & 26.02±2.00  & 27.39±2.12 & 28.99±2.28 &  {30.70±3.88}  & 29.89±2.32 & \textbf{31.05±3.04} \\
                                  &     & SSIM   & 58.37±9.63  & 74.87±5.28  & 77.86±6.11 & 81.82±5.98 & 75.88±13.73 & 84.06±7.67 & \textbf{85.82±8.58}    \\ \cmidrule(l){2-10} 
                                  & \multirow{2}{*}{15x} & PSNR &   24.95±1.82  & 20.37±0.97  & 17.79±1.27 & 17.52±2.50 & 29.99±3.69  &28.32±2.68 & \textbf{30.70±2.95}  \\
                                  &     & SSIM &  54.21±12.91 & 52.46±5.83  & 50.14±7.27 & 59.87±5.51 & 73.11±15.37 & 84.64±7.96 & \textbf{86.55±8.29}  \\ \midrule
\multirow{4}{*}{VD Poisson Disk}      & \multirow{2}{*}{8x} & PSNR &  24.42±1.94  & 25.91±2.12  & 27.81±2.31 & 26.24±2.07 & 30.96±4.01  & 30.29±3.01 & \textbf{31.32±3.09}  \\
                                  &     & SSIM &   50.13±10.84 & 73.55±5.68  & 78.31±5.51 & 75.36±5.59 & 78.26±12.51 & 84.31±8.08 & \textbf{85.43±8.67} \\ \cmidrule(l){2-10} 
                                  & \multirow{2}{*}{15x}  & PSNR &  24.28±1.86  & 25.91±2.12 & 27.47±2.27 & 26.29±2.07 & 30.14±3.79 & 29.65±2.90 & \textbf{30.92±3.01}   \\
                                  &     & SSIM &  48.90±11.37 & 71.90±5.69  & 76.42±6.59 & 75.38±5.58 & 73.89±14.77 & 84.40±8.21 & \textbf{85.82±8.58}  \\  \bottomrule
\end{tabular}
}
\label{Table 1: single-coil}
\end{table*}

\subsubsection{Single-coil study}
Table \ref{Table 1: single-coil} reports the quantitative results of four sampling patterns with multiple R's. Figure \ref{fig:main} displays the 8x reconstruction of a representative image with four masks. For In-D sampling (Uniform 1D) and the similar Gaussian 1D sampling, the SoTA supervised method, DuDoRNet, remains competitive at R=4; yet our method is comparable to it at R=8x and surpasses it at R=12x. For other OOD samplings, \colorbox{supervisedcolor}{ours.B} consistently outperforms others while the SoTA MRI diffusion method, DiffuseRecon is the second-best method, yet struggled with the intensity drifting issue of diffusion models\cite{lin2023flaw} as Score-MRI (Fig. \ref{fig:main}, 8x Gaussian 2D). Astonishingly, our method is the only one whose performance does not drop significantly with high R's.

\subsubsection{Multi-coil study}
Table \ref{multi-coil} lists the results on the multi-coil FastMRI dataset. \colorbox{supervisedcolor}{Ours.B} outperforms others while \colorbox{unsupervisedcolor}{ours.A} is the second-best method.
\begin{table}[H]
\caption{Multi-coil experiments on FastMRI. PSNR (dB) and SSIM (\%) are reported as mean±std. The \textbf{best} and \underline{second best} results are marked.}
\centering
\resizebox{0.45\textwidth}{!}{
\begin{tabular}{@{}ll|lllllll@{}}
\toprule
    &      & \colorbox{unsupervisedcolor}{TV} & \colorbox{unsupervisedcolor}{Ours.A} & \colorbox{supervisedcolor}{UNet}&
 \colorbox{supervisedcolor}{E2E-VarNet} & \colorbox{diffusioncolor}{Score-MRI} & \colorbox{diffusioncolor}{Ours.B}\\ \midrule
4x  & PSNR & 25.53±1.89  & 30.88±2.21  & 30.09±2.54 & 31.57±2.91 & \textbf{32.17±2.79} & \underline{31.88±2.73}     \\
    & SSIM & 69.63±13.50 & \underline{91.56±8.23}  & 89.00±6.41 & 89.63±5.82 & 88.18±4.17 & \textbf{92.46±8.45}\\ \midrule
8x  & PSNR & 24.15±2.08  & \underline{30.54±2.16} & 28.56±2.78 & 28.77±3.00 & 28.58±2.72 & \textbf{31.46±2.38}      \\
    & SSIM &  65.83±15.91 & \underline{91.20±8.25} & 86.41±6.01 & 86.94±6.43 & 85.66±5.26 & \textbf{92.16±8.32}    \\ \midrule
12x & PSNR &  23.87±1.88  & \underline{30.31±2.17} & 24.56±2.30 & 24.68±3.38 & 27.51±2.62 & \textbf{31.38±2.45}      \\
    & SSIM &  64.32±17.37 & \underline{90.45±8.48} & 82.10±6.64 & 82.43±6.96 & 83.82±5.62 & \textbf{91.72±8.60}    \\ \bottomrule
    
\end{tabular}
}
\label{multi-coil}
\end{table}

\subsection{Out-of-Distribution Contrast-and-Modality Study}
Table \ref{Table 2: ood modality} reports the generalizability of various methods trained on FastMRI towards combinations of OOD contrasts and OOD organ under Uniform 1D sampling with R=(4x,8x,12x).  \colorbox{unsupervisedcolor}{Ours.A} generalizes the best towards unseen MRI modalities and organs. As a reference of the upper bound, we also present the In-D performance of supervised methods, as well as ours. \colorbox{supervisedcolor}{Ours.B} shows similar behavior to that in the FastMRI studies, outperforming supervised methods in high (R=8x) and OOD (R=12x) acceleration rates.


\begin{table*}[]
\caption{Out-of-distribution contrasts-and-organ experiments on IXI. PSNR (dB) and SSIM (\%) are reported as mean±std. The \textbf{best} results are marked.}
\centering
\resizebox{0.8\textwidth}{!}{
\begin{tabular}{@{}c|cc|ccccc|ccc@{}}
\toprule
& & & & & & & & \multicolumn{3}{|l}{Reference: In-domain models' performance.} \\
OOD MRI Types    &   &     & \colorbox{unsupervisedcolor}{TV} & \colorbox{supervisedcolor}{UNet} & \colorbox{supervisedcolor}{DuDoRNet} & \colorbox{diffusioncolor}{Score-MRI}  & \colorbox{unsupervisedcolor}{Ours.A}  & \colorbox{supervisedcolor}{UNet} & \colorbox{supervisedcolor}{DuDoRNet}  &  \colorbox{diffusioncolor}{Ours.B}  \\ \midrule
\multirow{6}{*}{PD brain} & \multirow{2}{*}{4x}  & PSNR & 25.88±1.54 & 25.33±1.34 & 24.35±1.40 & 28.64±1.60 & \textbf{31.37±1.85} & 27.67±1.55 & \textbf{35.05±2.69} & 32.48±1.38 \\
                          &                      & SSIM &  74.86±2.68 & 78.93±2.87 & 72.29±4.10 & 87.64±1.93 & \textbf{94.06±1.67} & 87.59±2.48 & \textbf{97.41±1.08} & 95.45±3.35 \\ \cmidrule(l){2-11} 
                          & \multirow{2}{*}{8x}  & PSNR & 24.07±1.40 & 23.54±1.26 & 22.10±1.50 & 24.74±1.35 & \textbf{29.28±1.78} & 24.13±1.40 & 28.44±1.90 & \textbf{31.95±1.40} \\
                          &                      & SSIM & 71.35±2.78 & 77.61±3.14 & 70.83±4.72 & 79.96±2.21 & \textbf{89.77±2.11} & 85.27±2.39 & 92.40±2.18 & \textbf{94.75±3.47} \\ \cmidrule(l){2-11} 
                          & \multirow{2}{*}{12x} & PSNR & 24.02±1.38 & 23.34±1.34 & 21.66±1.52 & 24.32±1.31 & \textbf{28.08±1.79} & 23.70±1.40 & 26.84±1.52 & \textbf{31.17±1.40} \\
                          &                      & SSIM & 71.80±2.83 & 78.94±2.87 & 70.37±4.73 & 78.34±2.27 & \textbf{87.48±2.47} & 84.26±2.39 & 90.27±2.43 & \textbf{93.92±3.59} \\ \midrule
\multirow{6}{*}{T$_1$-w brain} & \multirow{2}{*}{4x}  & PSNR & 26.70±2.27 & 25.91±2.12 & 25.33±1.71 & 29.21±2.33 & \textbf{30.63±2.40} & 28.75±1.87 & \textbf{32.42±2.67} & 31.62±1.54\\ 
                          &                      & SSIM & 71.76±7.52 & 73.55±5.68 & 72.77±5.76 & 86.99±2.92 & \textbf{92.07±2.90} & 87.94±5.41 & 88.06±3.09  & \textbf{93.91±3.83}   \\ \cmidrule(l){2-11} 
                          & \multirow{2}{*}{8x}  & PSNR &  24.92±2.04 & 24.33±1.82 & 23.49±1.74 & 25.52±1.94 & \textbf{28.54±2.31} & 25.74±1.74 & 28.05±2.12 & \textbf{30.61±1.38}      \\
                          &                      & SSIM &  66.76±9.54 & 75.01±7.07 & 70.85±7.87 & 78.54±4.72 & \textbf{87.16±4.01} & 84.41±4.74 & 85.38±2.05  & \textbf{94.15±2.66}  \\ \cmidrule(l){2-11} 
                          & \multirow{2}{*}{12x} & PSNR &  24.83±2.06 & 24.18±1.84 & 22.98±1.85 & 25.03±1.94 & \textbf{27.32±2.22} & 25.14±1.80 & 26.87±2.12 & \textbf{30.10±1.55}  \\
                          &                      & SSIM &  67.11±9.71 & 74.94±7.30 & 70.15±9.04 & 76.66±4.96 & \textbf{84.18±4.74} & 83.22±4.73 & 85.13±5.09 & \textbf{91.67±4.16}   \\ \midrule
\multirow{6}{*}{T$_2$-w brain} & \multirow{2}{*}{4x}  & PSNR &   25.99±1.32 & 25.60±1.18 & 25.40±1.20 & 28.78±1.40 & \textbf{30.57±1.40} & 27.76±1.41 & \textbf{34.21±1.96} & 31.68±1.23 \\
                          &                      & SSIM & 75.79±2.53 & 80.54±2.44 & 77.31±3.55 & 88.61±1.52 & \textbf{93.69±1.60} & 87.97±2.18 & \textbf{97.25±1.02} & 94.94±3.44  \\ \cmidrule(l){2-11} 
                          & \multirow{2}{*}{8x}  & PSNR & 24.19±1.20 & 23.75±1.13 & 23.09±1.15 & 24.72±1.22 & \textbf{28.61±1.36} & 24.54±1.32 & 28.03±1.54 & \textbf{31.10±1.25}  \\
                          &                      & SSIM &  72.53±2.86 & 79.23±2.43 & 75.98±3.83 & 80.27±1.93 & \textbf{89.48±2.35} & 85.38±2.05 & 92.11±2.13 & \textbf{94.18±3.61}  \\ \cmidrule(l){2-11} 
                          & \multirow{2}{*}{12x} & PSNR &  24.11±1.19 & 24.18±1.84 & 22.75±1.15 & 24.32±1.18 & \textbf{27.08±1.31} & 23.91±1.28 & 25.33±1.35 & \textbf{30.30±1.24}   \\
                          &                      & SSIM & 72.82±2.89 & 74.94±7.30 & 76.63±2.80 & 78.62±2.02 & \textbf{86.53±2.52} & 83.73±2.11 & 86.91±2.13 & \textbf{93.38±3.72}  \\ 
\bottomrule
\end{tabular}
}
\label{Table 2: ood modality}
\end{table*}




\subsection{Universal Reconstruction with MRAdapter}
\label{sec:universal}
\begin{table*}[]
\caption{Universal reconstruction experiments. The LLDM's autoencoder is fine-tuned with single- and multiple-source datasets, listed in the first row. PSNR (dB) and SSIM (\%) are reported as mean±std. The \textbf{best} and \underline{second best} results are marked.}

\centering
\resizebox{0.8\textwidth}{!}{
\begin{tabular}{@{}c|cc|cccccc@{}}
\toprule
Recon. Type                                             &                               &      & None          & PD-w knee       & T$_1$-w brain     & T$_2$-w brain     & PD-w brain     & T$_1$+T$_2$+PD-w brain \\ \midrule
\multirow{8}{*}{Autoencoder's}     & \multicolumn{1}{l|}{PD-w knee}  & PSNR & 28.98 ± 3.10  & \textbf{31.62 ± 3.14}  & 31.27 ± 2.94 & 30.76 ± 2.65 & 30.58 ± 2.69 & \underline{31.17 ± 2.92}   \\
\multicolumn{1}{l|}{}                        & \multicolumn{1}{l|}{}         & SSIM & 79.17 ± 11.00 & \textbf{87.02 ± 8.24}  & 86.32 ± 8.30 & 84.44 ± 8.27 & 83.44 ± 8.09 & \underline{85.61 ± 8.21}   \\ \cmidrule(l){2-9} 
\multicolumn{1}{l|}{}                        & \multicolumn{1}{l|}{T$_1$-w brain} & PSNR & 28.32 ± 2.33  & 30.23 ± 2.38  & \underline{30.68 ± 5.38} & 30.27 ± 2.32 & 30.28 ± 2.25 & \textbf{30.71 ± 2.32}   \\
\multicolumn{1}{l|}{}                        & \multicolumn{1}{l|}{}         & SSIM & 90.37 ± 3.46  & 93.12 ± 2.60  & \underline{93.43 ± 2.39} & 92.94 ± 2.54 & 92.79 ± 2.45 & \textbf{93.50 ± 2.42}   \\ \cmidrule(l){2-9} 
\multicolumn{1}{l|}{}                        & \multicolumn{1}{l|}{T$_2$-w brain} & PSNR & 27.98 ± 1.43  & 29.77 ± 1.42  & 29.98 ± 1.95 & \underline{30.34 ± 2.16} & 29.96 ± 1.99 & \textbf{30.42 ± 1.46}   \\
\multicolumn{1}{l|}{}                        & \multicolumn{1}{l|}{}         & SSIM & 91.79 ± 2.18  & 93.22 ± 1.65  & 93.41 ± 1.60 & \underline{94.36 ± 1.61} & 93.99 ± 1.65 & \textbf{94.56 ± 2.44}   \\ \cmidrule(l){2-9} 
\multicolumn{1}{l|}{}                        & \multicolumn{1}{l|}{PD-w brain} & PSNR & 28.89 ± 1.88  & 30.70 ± 1.90  & 30.96 ± 1.88 & 30.90 ± 1.90 & \underline{31.28 ± 1.96} & \textbf{31.41 ± 1.96}   \\
\multicolumn{1}{l|}{}                        & \multicolumn{1}{l|}{}         & SSIM & 92.73 ± 2.01  & 94.35 ± 1.76  & 94.02 ± 1.82 & 94.27 ± 1.79 & \underline{94.92 ± 1.50} & \textbf{95.18 ± 1.48}   \\ \midrule
\multirow{8}{*}{8x undersampled} & \multicolumn{1}{l|}{PD-w knee}  & PSNR & 30.26 ± 1.63  & \textbf{31.28 ± 1.76}  & 31.22 ± 3.04 & 31.18 ± 3.01 & 31.14 ± 3.01 & \underline{31.23 ± 3.06}   \\
\multicolumn{1}{l|}{}                        & \multicolumn{1}{l|}{}         & SSIM & 83.97 ± 8.41 & \textbf{85.98 ± 2.92}  & \underline{85.87 ± 8.56} & 85.79 ± 6.57 & 85.81 ± 8.59 & \underline{85.87 ± 8.59}   \\ \cmidrule(l){2-9} 
\multicolumn{1}{l|}{}                        & \multicolumn{1}{l|}{T$_1$-w brain} & PSNR & 28.54 ± 2.31  & 30.56 ± 1.51  & \textbf{31.07 ± 2.36} 
& \underline{30.87 ± 2.38} & 30.79 ± 2.32 & 30.61 ± 1.38   \\
\multicolumn{1}{l|}{}                        & \multicolumn{1}{l|}{}         & SSIM & 87.16 ± 4.01  & 92.47 ± 16.05 & \underline{93.22 ± 2.38} & 92.83 ± 2.52 & 92.86 ± 2.54 & \textbf{94.15 ± 2.66}   \\ \cmidrule(l){2-9} 
\multicolumn{1}{l|}{}                        & \multicolumn{1}{l|}{T$_2$-w brain} & PSNR & 28.61 ± 1.36  & 30.42 ± 1.41  & 30.90 ± 1.46 & \underline{31.09 ± 1.50} & 30.73 ± 1.45 & \textbf{31.10 ± 1.25}   \\
\multicolumn{1}{l|}{}                        & \multicolumn{1}{l|}{}         & SSIM & 89.48 ± 2.35  & 93.56 ± 12.86 & 94.16 ± 1.56 & \textbf{94.34 ± 1.60} & 94.06 ± 1.62 & \underline{94.18 ± 3.61} \\ \cmidrule(l){2-9} 
\multicolumn{1}{l|}{}                        & \multicolumn{1}{l|}{PD-w brain} & PSNR & 29.28 ± 1.78  & 31.25 ± 1.80  & 31.82 ± 1.86 & 30.73 ± 1.45 & \underline{31.89 ± 1.87} & \textbf{31.95 ± 1.40}   \\
\multicolumn{1}{l|}{}                        & \multicolumn{1}{l|}{}         & SSIM & 89.77 ± 2.11  & 94.15 ± 12.16 & 94.82 ± 1.38 & 94.06 ± 1.62 & \textbf{94.99 ± 1.40} & \underline{94.75 ± 3.47}  \\ \midrule
 
\multicolumn{3}{l|}{Number of slices in each training set} & 0 & 25012 & 3218 & 2455 & 2455 & 8128 \\\bottomrule
\end{tabular}
}
\label{Table 3: universal reconstruction}
\end{table*}

Table \ref{Table 3: universal reconstruction} summarizes the quantitative results of finetuning with five sets of single- and multi-source MRI on every separate dataset. Autoencoder's reconstruction and 8x uniform 1D reconstruction ability are displayed in two sections. Direct conclusions from the table are: (1) With single- or multi-source MRI datasets, VQGAN's MRI reconstruction ability is improved on every separate dataset, no matter whether In-D or OOD (section autoencoder's, column None versus the other columns). (2) VQGAN's reconstruction ability is boosted more with MRI datasets closer to testing sets (section autoencoder's, the diagonal of the single-source columns) or larger datasets  (section autoencoder's, PD knee and T$_1$-w+T$_2$-w+PD-w columns versus others). (3) With single- or multiple-source MRI data, our undersampled reconstruction ability is improved on all tested MRI datasets, no matter In-D or OOD (section 8x undersampled, column None versus the others). (4) Our undersampled reconstruction ability is boosted more with MRI datasets closer to testing sets (section 8x undersampled, the diagonal of the single-source columns) or larger datasets (section 8x undersampled, PD knee and T$_1$-w+T$_2$-w+PD-w columns versus others).

Combining two sections of Table \ref{Table 3: universal reconstruction}, we observe a counter-intuitive fact: the reconstruction of the ground truth MRI data is similar to or even worse than that of 8$\times$ undersampled MRI (differences between the same rows in two sections). It implies that lossy as VQGAN is, both our MRSampler (column None) and MRAdapter (all columns except None) can compensate for the information loss. An explanation is that MRSampler's soft classifier guidance term corrects the reconstruction result on the pixel domain directly so that the gradients guide the LLDM to generate pixel-domain consistent results.

\subsection{Hyperparameter Selection of MRPD}
We show the selection process on the single-coil FastMRI dataset with a 4x uniform 1D sampling mask.
\subsubsection{$\lambda$}
\label{lambda}
Table \ref{MRSampler's coefficient} shows quantitative results of different $\lambda$'s. The optimal $\lambda=1$ remains consistent among our experiments' different sampling patterns, modalities, and organs. This corresponds to discarding the original image phase.
\begin{table}[H]
    \vspace{-10pt}
    \centering
    \caption{Ablation study on MRSampler's $\lambda$.}
    \resizebox{0.45\textwidth}{!}{
        \begin{tabular}{@{}l|cccccccc@{}}
        \toprule
        $\lambda$  & 0 & 0.1 & 0.2 & 0.4 & 0.6 & 0.8 & 0.9 & 1 \\ \midrule
        PSNR (dB) & 27.99  &  29.12 & 29.65 & 29.92 & 29.42 & 30.68 & 31.17 & \textbf{31.50}\\
        SSIM (\%) & 81.05  & 83.13 & 83.13 & 82.83 & 83.95 & 85.58 & 86.21 & \textbf{86.60} \\ \bottomrule
        \end{tabular}
    }
    \label{MRSampler's coefficient}
    \vspace{-8pt}
\end{table}
\subsubsection{$t_0$,$t_{ws}$}
\label{t0,tws}
\begin{figure}[H]
    \vspace{-10pt}
    \centering
    \includegraphics[width=0.45\textwidth]{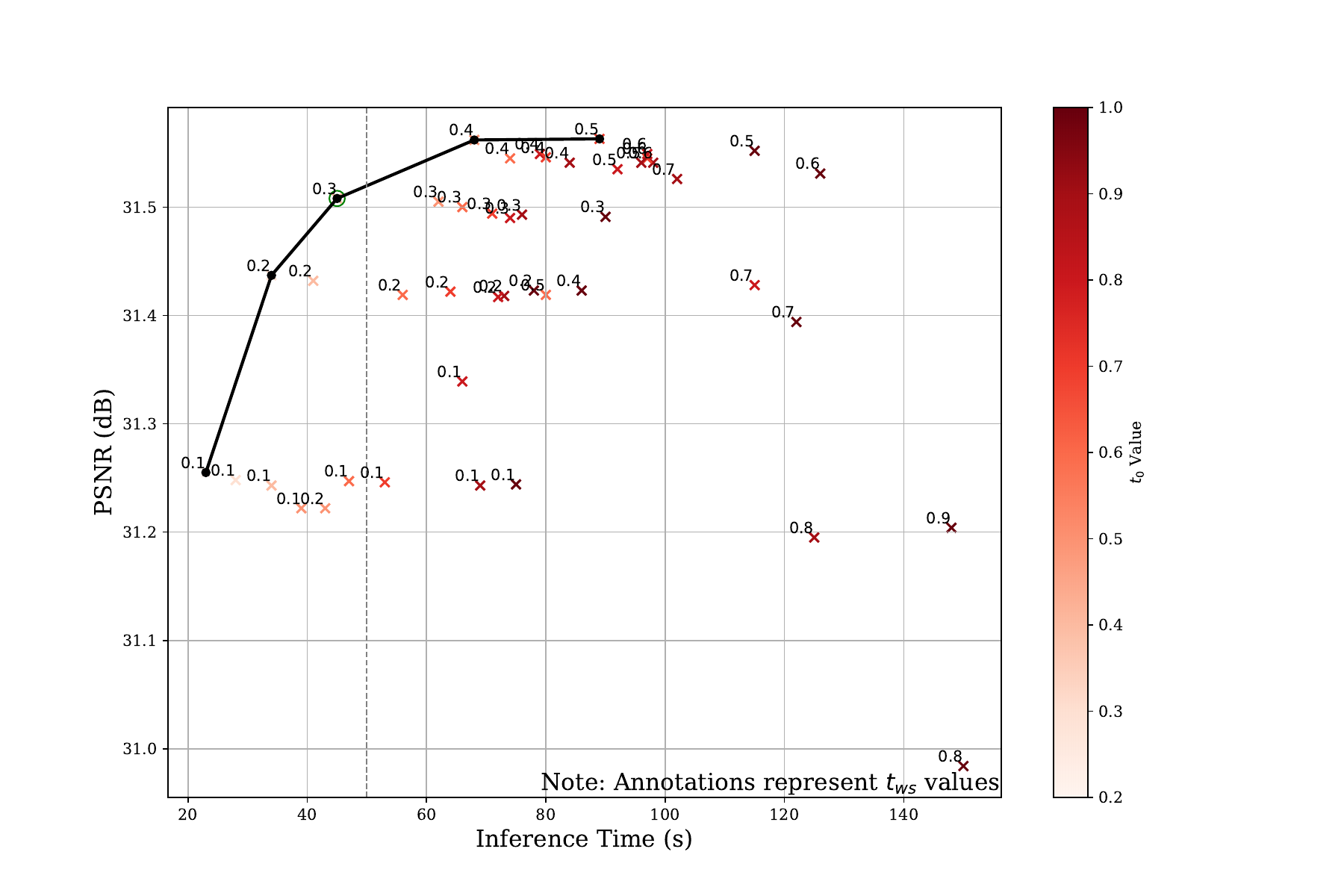}
    \caption{Pareto Frontier of inference time and PSNR under different combinations of $t_0$ and $t_{ws}$. The $\times$ markers represent feasible choices within the range of $t_0$,$t_{ws}$. The black line is the Pareto-efficient frontier and the gray line is a time budget of 50s. Our chosen value is \textcolor{mplgreen}{circled}.} 
    \label{fig:time-performance trade-off} 
\end{figure}

Figure \ref{fig:time-performance trade-off} shows the inference time and PSNR with different combinations of ($t_0$,$t_{ws}$). Our selected (0.4,0.3) is on the Pareto frontier and is the best choice given our inference time budget of 50s, DiffuseRecon\cite{peng2022DiffuseRecon}'s inference time. The optimal $t_0$,$t_{ws}$ remains consistent among different sampling patterns, modalities, and organs in our experiments.


\subsection{Ablation Studies}

\subsubsection{MRSampler design}
\begin{figure}[H]
    \vspace{-0.1in}
    \centering
    \includegraphics[width=0.45\textwidth]{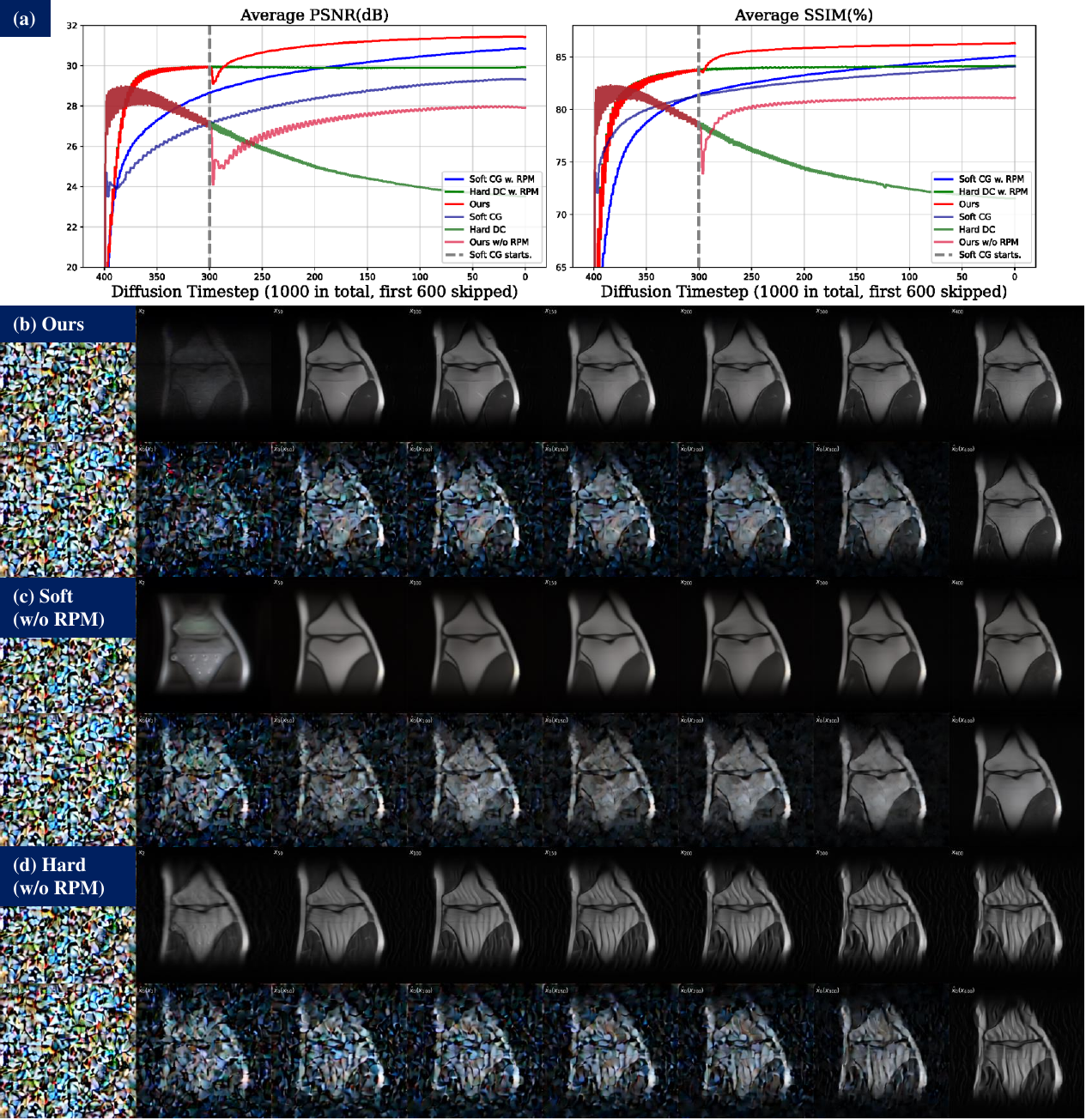}
    \caption{Image reconstruction evolution during the diffusion process guided by the MRSampler and other unitary or magnitude-only controllers. (a) The evolution of PSNR and SSIM. (b)-(d) In each panel, the first row illustrates the evolution of clean image predictions \(\hat{x}_0(z_t)\), and the second row depicts the intermediate noisy images \(z_t\).}
    \label{ResultOfRSampler}
    \vspace{-5pt}
\end{figure}

Figure \ref{ResultOfRSampler} displays the reconstruction process using MRSampler and various control mechanisms. Hard-to-soft control allows MRSampler to bypass local minima associated with hard consistency terms, resulting in faster convergence and superior performance compared to soft consistency terms alone. RPM improves the performance of both soft and hard control.

\subsubsection{MRAdapter architecture} Table \ref{ab:MRAdapter} shows the results of selective tuning of the VQGAN, including tuning the encoder and decoder components separately, as well as unadapted (Ori. VQGAN) and both-adapted (ours). The results are from the single-coil FastMRI dataset with R=4x,8x. While both tuning strategies—tuning the encoder alone and tuning the decoder alone—yield improvements in PSNR and SSIM over the original VQGAN, our integrated MRAdapter configuration achieves the best performance. 

\begin{table}[H]
    \caption{Ablation study on the MRAdapter. PSNR (dB) and SSIM (\%) are reported as mean±std.}
    \centering
        \vspace{-5pt}
    \resizebox{0.43\textwidth}{!}{
    \begin{tabular}{l|cc|cc}
    \toprule
    & \multicolumn{2}{c|}{\textbf{4x}} & \multicolumn{2}{c}{\textbf{8x}} \\
    \cmidrule(l){2-5} 
   \textbf{Architecture}  & PSNR \scriptsize{(dB)} & SSIM \scriptsize{(\%)}& PSNR \scriptsize{(dB)} & SSIM \scriptsize{(\%)}  \\ \midrule
    Original VQGAN & 30.88±1.63 & 85.62±8.23 & 30.26±1.63 & 83.97±8.41 \\
    Tune encoder & 31.35±1.69 & 86.59±2.88 & 31.23±1.74 & \textbf{86.09±2.92} \\
    Tune decoder & 31.31± 1.68 & 86.39±2.90 & 30.85±1.68 & 85.22±2.91 \\
    Ours & \textbf{31.50±1.74} & \textbf{86.60±2.89} & \textbf{31.28±2.91} & 85.98±8.56 \\
    \bottomrule
    \end{tabular}
    }
\label{ab:MRAdapter}
\end{table}

\section{Discussion}

\subsection{Time-Performance Trade-off}
Besides the performance, times for training and inference are two practical concerns for deploying MRI reconstruction algorithms. Table \ref{tab:time} lists the computation time on single-coil FastMRI. Note that \colorbox{unsupervisedcolor}{TV} and \colorbox{unsupervisedcolor}{ours.A} have no training overhead. Our model, built on the foundation of LLDMs, has the shortest training time (0/0.5 day) while maintaining sampling-agnostic. During inference, supervised models are faster than existing diffusion models, including ours. However, we believe our additional inference time is worth the gains.
\begin{table}[H]
    \centering
    \vspace{-10pt}
    \caption{Training time per dataset and inference time per slice for reconstructions at R=4x in FastMRI. }
    \resizebox{0.48\textwidth}{!}{
    \begin{tabular}{c|ccccccc}
      \toprule
         &  \colorbox{unsupervisedcolor}{TV} & \colorbox{supervisedcolor}{UNet} & \colorbox{supervisedcolor}{DuDoRNet} & \colorbox{diffusioncolor}{DiffuseRecon} & \colorbox{diffusioncolor}{Score-MRI} & \colorbox{unsupervisedcolor}{Ours.A} & \colorbox{diffusioncolor}{Ours.B}\\
      \midrule
      Training & - & 2.5 days & 7 days & 9 days & 3 weeks & - & 0.5 day \\
      Inference & 2s & 0.04s & 0.13s & 50s & 120s/18s(w. CCDF\cite{chung2022ccdf})& 45s & 45s \\ \bottomrule
    \end{tabular}}
    \label{tab:time}
        \vspace{-5pt}
\end{table}
Firstly, conventional supervised models require training and deployment for each specific imaging scenario, limiting their clinical application in the real world\cite{cmrxrecon2024}. As a result, generalizable universal models\cite{ouyang2019generalizing,wang2023onegeneralizable,dar2020transfergeneralizable,liu2021universal} are highly desirable for efficient deployment on MRI machines. Among all compared methods, MRPD generalizes the best towards different sampling patterns, organs, and contrasts and improves the performance universally with finetuning. Secondly, acquiring fully sampled datasets is infeasible in many practical imaging scenarios\cite{yaman2020ssdu}, including real-time imaging and intraoperative MRI\cite{mislow2010originsiMRI}, where highly undersampled images are acquired. Even available, aggregating heterogeneous realistic MRI data from various hospitals is infeasible for data privacy issues\cite{wang2023onegeneralizable}. MRPD suits well for those real-world scenarios with both database-free (ours.A) and database-available (ours.B) versions and the superior performance at high R's. Lastly, it is viable to accelerate MRPD with off-the-shelf diffusion acceleration techniques. Table \ref{DDIM accelerations} shows the performance and inference time of our methods with DDIM acceleration\cite{song2021DDIM} on single-coil FastMRI (R=4$\times$). At T=500, our method is accelerated for $2\times$ while maintaining a decent performance. It is also promising to leverage SoTA MRI diffusion acceleration techniques\cite{chung2022ccdf,ozturkler2023red-diff,Cao2024HFS-SDE} for a faster inference.

\begin{table}[H]
    \vspace{-10pt}
    \centering
    \caption{Our method with the DDIM acceleration\cite{song2021DDIM}.}
    \resizebox{0.48\textwidth}{!}{
    \begin{tabular}{@{}c|ccc@{}}
    \toprule
    Total DDIM Step (T)     & 100 & 500 & 1000  \\ \midrule
    PSNR (dB) / SSIM (\%) &  28.52 / 80.70 & 31.21 / 86.08  &  31.50 / 86.60\\
    Inference Time (s) & 8 & 25 & 45\\
    \bottomrule
    \end{tabular}}
    \label{DDIM accelerations}
    \vspace{-5pt}
\end{table}
\subsection{The Rationale of RPM}
Similar to DiffuseRecon\cite{peng2022DiffuseRecon} and Resample\cite{song2023resample}, we discover the advantage of introducing stochasticity to the intermediate images for reconstruction and propose RPM to inject a separable noise to MR images by combining the original image phase with a random one. In Table \ref{lambda}, we find that the more original phase is added, the worse the reconstruction is. Meanwhile, Fig. \ref{ResultOfRSampler}(a) visualizes our method w/o RPM, corresponding to $\lambda=0$. PSNR and SSIM drop step by step, meaning that unmodulated images deviate from the correct data manifold during sampling. This indicates that given the strong LLDM prior, the benefits of proper noise injection significantly outweigh the drawback of discarding the original undersampled image phase for MRI reconstruction. 

\subsection{Limitations and Future Work}
Current MRPD requires a longer inference time than supervised models. To speed up, we have exploited the utilization of skipping the early diffusion steps \cite{meng2022sdedit} and enforcing hard DC and soft CG every two steps \cite{song2023resample}. In the previous section, we show the results of DDIM acceleration. These preliminary results demonstrate the potential of MRPD to be further accelerated. Furthermore, parallel computing using multiple GPUs for multi-coil acquisition is viable for MRPD. 

MRPD is a proof-of-concept to utilize LLDMs for MRI reconstruction and can be combined with existing test-time adaptation methods\cite{ozturkler2023smrd} for further improvements.
\vspace{-4pt}

\section{Conclusion}
We have successfully demonstrated that an LLDM contains visual knowledge that is universal for natural and medical images and can be leveraged to improve the adaptability and generalizability of undersampled MRI reconstruction. We propose MRPD, a novel unsupervised and universally adapted framework based on an LLDM pre-trained on vast natural images. We validate through rigorous tests, encompassing a variety of sampling patterns, MRI contrasts, and organs, and have yet to find a single data type on which our MRPD does not generalize. Furthermore, MRPD adapts to scenarios both with and without access to MRI databases. The superior generalizability and the universally adaptive characteristics make MRPD potentially catalyze related foundation-model-based research for accelerated MRI reconstruction.

\bibliography{tmi}
\bibliographystyle{ieeetr}

\end{document}